\documentclass[sigconf, nonacm, anonymous=false, pdfa]{acmart}

\usepackage{listings}
\usepackage{siunitx}
\usepackage{multirow}
\usepackage{hyperref}
\PassOptionsToPackage{capitalize,noabbrev,nameinlink}{cleveref}
\usepackage{cleveref}
\usepackage{xspace}
\usepackage{enumitem}
\usepackage[outline]{contour}
\usepackage{xcolor}
\usepackage{colortbl}
\usepackage{float}
\usepackage{caption}
\usepackage{subcaption}
\usepackage{multicol}
\usepackage{titlesec}

\usepackage{nameref}

\newcommand{\Pararef}[1]{\hyperref[#1]{Paragraph~\nameref*{#1}}}
\newcommand{\pararef}[1]{\hyperref[#1]{Paragraph~\nameref*{#1}}}

\usepackage{tikz}
\usetikzlibrary{decorations.pathmorphing}
\usepackage[a-2b,mathxmp]{pdfx}[2018/12/22]

\titlespacing*{\paragraph}{0pt}{0pt}{1ex}

\titleformat{\paragraph}[runin]
  {\normalfont\normalsize\bfseries}  
  {}                                 
  {0pt}                              
  {}                                 

\titleformat{\section}{\normalfont\Large\bfseries}{\thesection}{1em}{\MakeTextUppercase}

\newfloat{lstfloat}{htbp}{lop}
\floatname{lstfloat}{Listing}

\hypersetup{
     colorlinks=true,
     linkcolor=blue,
     filecolor=blue,
     citecolor = black,
     urlcolor=cyan,
    }

\lstdefinestyle{cpp} {
  language = C++,
  basicstyle = {\ttfamily \footnotesize},
  keywordstyle = {\color{blue!40!blue}},
  keywordstyle = [2]{\color{blue!60!black}},
  morekeywords = [2]{uint64_t, uint32_t, uint16_t, uint8_t, int64_t, int32_t, int16_t, \
    int8_t, int, float, double, char, void, __m512i, __mmask16, __mmask8},
} 

\newcommand{\iouring}{io\texttt{\_}uring}

\DeclareRobustCommand{\ttsafe}[1]{%
  \begingroup
    \ttfamily
    \detokenize{#1}%
  \endgroup
}

\graphicspath{{figures/}}

\newcommand\vldbdoi{10.14778/3819518.3819553}
\newcommand\vldbpages{2317-2330}
\newcommand\vldbvolume{19}
\newcommand\vldbissue{9}
\newcommand\vldbyear{2026}
\newcommand\vldbauthors{\authors}
\newcommand\vldbtitle{\shorttitle} 
\newcommand\vldbavailabilityurl{https://github.com/mjasny/vldb26-iouring}
\newcommand\vldbpagestyle{empty}

\newcommand{\hl}[1]{\textcolor{black}{#1}}
\newenvironment{highlight}
{\color{black} \captionsetup{font={color=black}}}
{}

\begin{document}
\title{High-Performance DBMSs with \iouring{}: When and How to Use It}



\author{Matthias Jasny}
\affiliation{
  \institution{Technische Universität Darmstadt}
  \city{}\country{}
}
\email{matthias.jasny@tu-darmstadt.de} 
\author{Muhammad El-Hindi}
\affiliation{
  \institution{Technische Universität München}
  \city{}\country{}
}
\email{muhammad.el-hindi@tum.de}
\author{Tobias Ziegler}
\affiliation{
  \institution{TigerBeetle}
  \city{}\country{}
}
\email{tobias@tigerbeetle.com}
\author{Viktor Leis}
\affiliation{
  \institution{Technische Universität München}
  \city{}\country{}
}
\email{viktor.leis@tum.de}
\author{Carsten Binnig}
\affiliation{
  \institution{Technische Universität Darmstadt \&\\ DFKI \&  hessian.AI}
  \city{}\country{}
}
\email{carsten.binnig@tu-darmstadt.de}

\begin{abstract}
We study how modern database systems can leverage the Linux \iouring{} interface for efficient, low-overhead I/O.
\iouring{} is an asynchronous system call batching interface that unifies storage and network operations, addressing limitations of existing Linux I/O interfaces.
However, naively replacing traditional I/O interfaces with \iouring{} does not necessarily yield performance benefits.
To demonstrate when \iouring{} delivers the greatest benefits and how to use it effectively in modern database systems, we evaluate it in two use cases:
Integrating \iouring{} into a storage-bound buffer manager and using it for high-throughput data shuffling in network-bound analytical workloads.
We further analyze how advanced \iouring{} features, such as registered buffers and passthrough I/O, affect end-to-end performance.
Our study shows when low-level optimizations translate into tangible system-wide gains and how architectural choices influence these benefits.
Building on these insights, we derive practical guidelines for designing I/O-intensive systems using \iouring{} and validate their effectiveness in a case study of PostgreSQL's recent \iouring{} integration, where applying our guidelines yields a performance improvement of 14\%.
\end{abstract}

\maketitle

\pagestyle{\vldbpagestyle}
\begingroup\small\noindent\raggedright\textbf{PVLDB Reference Format:}\\
\vldbauthors. \vldbtitle. PVLDB, \vldbvolume(\vldbissue): \vldbpages, \vldbyear.\\
\href{https://doi.org/\vldbdoi}{doi:\vldbdoi}
\endgroup
\begingroup
\renewcommand\thefootnote{}\footnote{\noindent
This work is licensed under the Creative Commons BY-NC-ND 4.0 International License. Visit \url{https://creativecommons.org/licenses/by-nc-nd/4.0/} to view a copy of this license. For any use beyond those covered by this license, obtain permission by emailing \href{mailto:info@vldb.org}{info@vldb.org}. Copyright is held by the owner/author(s). Publication rights licensed to the VLDB Endowment. \\
\raggedright Proceedings of the VLDB Endowment, Vol. \vldbvolume, No. \vldbissue\ %
ISSN 2150-8097. \\
\href{https://doi.org/\vldbdoi}{doi:\vldbdoi} \\
}\addtocounter{footnote}{-1}\endgroup

\ifdefempty{\vldbavailabilityurl}{}{
\vspace{.3cm}
\begingroup\small\noindent\raggedright\textbf{PVLDB Artifact Availability:}\\
The source code, data, and/or other artifacts have been made available at \url{https://github.com/mjasny/vldb26-iouring}.
\endgroup
}

\section{Introduction}
\label{sec:introduction}

\begin{figure}[]
  \centering
  \captionsetup{aboveskip=0.0ex,belowskip=0.0ex}
  \captionsetup[subfigure]{aboveskip=0.0ex,belowskip=0.0ex}
  \includegraphics[width=0.99\linewidth]{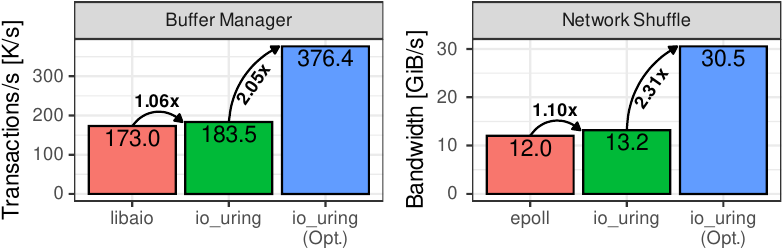}
  \caption{Performance comparison between traditional I/O interfaces and \iouring{} in a buffer manager and network shuffle. Naive use only yields modest gains, whereas designs that fully exploit \iouring{} more than double the performance.}
  \label{fig:intro_plot}
\end{figure}

\paragraph{Modern hardware and the I/O bottleneck.}
Modern PCIe 5.0 hardware, including NVMe SSDs such as the Kioxia CM7-R (2.45M IOPS) and NICs such as the ConnectX-7 (400 Gbit/s), sustains millions of IOPS and hundreds of gigabits per second of throughput, yet conventional I/O interfaces struggle to saturate them~\cite{connectx7, DBLP:journals/pvldb/WuCCIS25,DBLP:journals/pvldb/HeXLJZYML23,DBLP:journals/pvldb/HaasL23,DBLP:journals/pvldb/LeisD24}.
In particular, kernel-based I/O interfaces, as still used by production-grade database systems, incur system-call and context-switch overhead, consuming a significant fraction of CPU cycles without saturating these devices~\cite{DBLP:conf/cidr/ZhouLYS25}.
These inefficiencies widen the hardware--software gap and make low-overhead I/O mechanisms central to modern database systems.


\paragraph{Challenges of user-space I/O.}
User-space I/O frameworks such as DPDK, SPDK, and RDMA bypass the kernel and can deliver high performance on dedicated hardware~\cite{DBLP:conf/sigmod/LiDSN16,DBLP:conf/cidr/ZhouLYS25,DBLP:conf/damon/JasnyE0B25}.  
However, operating entirely in user space removes OS abstractions such as file systems and TCP networking, making integration difficult for production databases that rely on them.  
These stacks also require exclusive control of SSDs or NICs~\cite{DBLP:journals/pvldb/ButrovichRRLZSP23,DBLP:conf/nsdi/FriedCSCGEFB24,DBLP:conf/sigcomm/KimMBZLPRSSS18}, which may conflict with deployments where devices must be shared.  
Consequently, user-space I/O, despite its advantages, has not seen wide adoption and is used mainly in specialized, tightly controlled environments rather than general-purpose systems~\cite{DBLP:journals/pvldb/ButrovichRRLZSP23,DBLP:conf/cidr/HaasHL20}.

\paragraph{\iouring{} features for efficient I/O.}
The Linux \iouring{} interface~\cite{io_uring} is a promising candidate to bridge the gap between efficient I/O and the preservation of common kernel abstractions.
It combines three key features, distinguishing it from earlier kernel I/O interfaces.
First, \emph{a unified interface} integrates storage, network, and other system calls into one framework.
\hl{Second, \emph{fully asynchronous execution} overcomes limitations of existing interfaces, allowing applications to perform useful work while I/O operations complete in the background.}
Third, \emph{batched submission and completion} process multiple operations with a single system call, amortizing system call overhead and context switches.
These features make \iouring{} attractive for database systems that issue large numbers of storage and network I/O operations.

\paragraph{Low-overhead I/O with \iouring{}?}
However, \iouring{} is not a panacea.
Simply replacing traditional I/O interfaces with \iouring{} does not necessarily yield substantial performance benefits.
As \Cref{fig:intro_plot} shows, using \iouring{} off the shelf instead of \texttt{libaio} for storage I/O in a buffer manager, or instead of \texttt{epoll} for a network shuffle, only modestly improves performance  (by 1.06$\times$ and 1.10$\times$, respectively).
In contrast, when the system is explicitly designed around \iouring{}'s capabilities (e.g., batching) and uses appropriate optimizations (e.g., registered buffers), the end-to-end performance improvements become much more pronounced:  2.05$\times$ for the buffer manager and 2.31$\times$ for the network shuffle.

These observations motivate our three research questions to guide DBMS system builders in using \iouring{}:
\begin{enumerate}[leftmargin=*,nosep]
    \item \emph{When to use \iouring{}?} Under which conditions -- especially I/O-intensive scenarios -- does it provide the greatest benefit?
    \item \emph{How to integrate \iouring{}?} How should a DBMS architecture incorporate \iouring{} to exploit its capabilities effectively?
    \item \emph{How to tune \iouring{}?} Which \iouring{} features most strongly influence DBMS performance?
\end{enumerate}

\paragraph{Contributions and outline.}
\begin{highlight}
In this paper, we study \iouring{} as a kernel-based path to high-performance I/O that preserves standard OS abstractions in database systems, across both storage- and network-bound workloads.
\end{highlight}
We evaluate \iouring{} using two complementary use cases supplemented with microbenchmarks.
First, we integrate \iouring{} into a transactional storage engine on NVMe SSDs to examine storage-bound workloads (\Cref{sec:storage-intensive}).
Second, we employ \iouring{} for data shuffling in a distributed analytics engine on 400~Gbit/s networks, representing network-bound workloads (\Cref{sec:network-intensive_systems}).
From these case studies, we derive general principles for effective \iouring{} use and validate them by improving PostgreSQL's \iouring{} backend to achieve more than 10\% additional speedup (\Cref{sec:insights}).

\section{Background on \iouring{}}
\label{sec:background}

\begin{figure}
    \centering
    \captionsetup{aboveskip=0.0ex,belowskip=0.0ex}
    \captionsetup[subfigure]{aboveskip=0.0ex,belowskip=0.0ex}
    \includegraphics[width=0.31\textwidth]{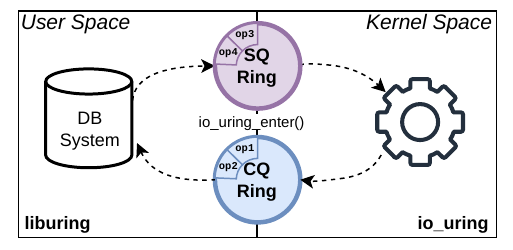}
    \caption{\iouring{} architecture. The database system in user space communicates with the kernel via two shared ring buffers: the Submission Queue (SQ) for enqueuing I/O requests and the Completion Queue (CQ) for receiving results.}
    \label{fig:basic_uring}
\end{figure}

\iouring{} was introduced into the Linux kernel in 2019 and has since been actively developed and optimized. 
Despite this progress, there has been little work on understanding how to adapt it to the requirements of database systems.
This section therefore provides the background needed to understand how \iouring{} can be used in DBMSs and how its design influences performance.
We highlight two aspects that distinguish \iouring{} from existing I/O backends: its application interface and its internal execution model, both of which enable high performance for data-intensive systems.

\subsection{Interface of \iouring{}}
Through its three key capabilities (unified I/O, asynchronous execution, and batching), \iouring{} provides an interface that aligns well with the demands of high-speed DBMSs.
Below, we describe these capabilities in more detail and their implications for DBMSs.

\paragraph{Unification of I/O with \iouring{}.}
Traditional DBMSs rely on synchronous system calls such as \texttt{read()} and \texttt{write()}, which provide a simple, uniform abstraction but scale poorly due to their blocking behavior.
To enable non-blocking network I/O, Linux introduced \texttt{epoll} to monitor multiple sockets for readiness.
For storage, \texttt{libaio} provided a separate API that in practice was mostly restricted to direct block I/O and often lacked true asynchronous execution.
This fragmentation forced developers to combine \texttt{epoll} and \texttt{libaio}, leading to duplicated code paths and limited concurrency between storage and network operations.
\iouring{} eliminates the need for such hybrid designs by unifying storage and network (as well as other system calls) under a single fully asynchronous API.
Its capabilities continue to expand beyond I/O-related system calls (e.g., \texttt{madvise}), moving toward a general-purpose asynchronous execution model for Linux.
The unified interface enables DBMSs to overlap network and disk I/O more efficiently within a single path, simplifying system design and reducing context switching.

\begin{figure}
    \centering
    \captionsetup{aboveskip=0.0ex,belowskip=0.0ex}
    \captionsetup[subfigure]{aboveskip=0.0ex,belowskip=0.0ex}
    \includegraphics[width=0.79\linewidth]{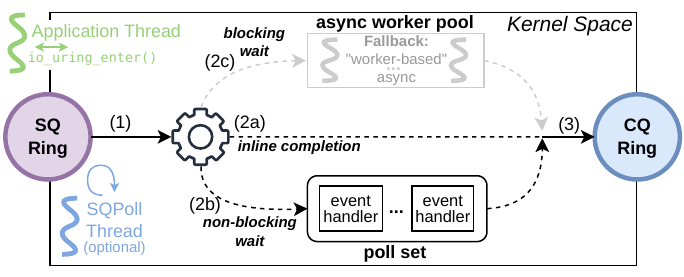}
    \caption{\iouring{} provides three execution paths: inline completion (2a), asynchronous execution via the poll set (2b), and a fallback to worker threads for blocking operations (2c).
    An optional SQPoll thread can submit requests without syscalls.}
  \label{fig:iouring_work}
\end{figure}

\paragraph{Asynchronous architecture.}
At a high level, \iouring{} is as a unified asynchronous layer over existing kernel I/O subsystems, such as the block layer for storage devices and the TCP/IP stack for networking.
It is implemented in the kernel (\Cref{fig:basic_uring}, right) and accessed through the \texttt{liburing} user-space library~\cite{liburing}.
Unlike \texttt{epoll}'s readiness-based polling approach, \iouring{} employs a \emph{completion-based} model, notifying applications after operations complete rather than when they become possible.
It implements this model using two memory-mapped ring buffers: the \emph{Submission Queue (SQ)} and the \emph{Completion Queue (CQ)}. 
These queues are shared between user space and the kernel, avoiding additional data copies when submitting and completing requests.
Applications enqueue I/O requests in the SQ, and their corresponding completions later appear in the CQ.
Because completions may arrive out of order, each request carries a user-defined identifier to match submissions and completions.
\iouring{} further supports \emph{request linking} to enforce operation ordering for multiple elements in the SQ.

\paragraph{Batch processing.}
While \texttt{epoll} can report multiple readiness events, each I/O operation, such as \texttt{read()}, still requires its own syscall.
In contrast, \iouring{} enables applications to enqueue multiple I/O requests in the SQ before triggering their submission with a single \texttt{io\_uring\_enter} syscall.
Similarly, multiple completions can be retrieved from the CQ in one step.
This batching capability amortizes syscall overhead and reduces context switches, allowing the kernel to process operations in bulk.
Even modest batch sizes (e.g., 16 operations) reduce the CPU cycles per operation by roughly 5--6$\times$ compared to single-operation submission as shown below:

\noindent
\begin{minipage}{\linewidth}
    \vspace{0.5em}
    \centering
    \includegraphics[width=0.60\linewidth]{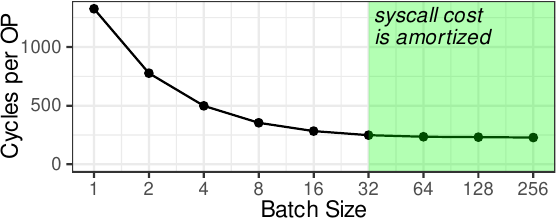}
    \label{fig:batching}
    \vspace{0.5em}
\end{minipage}

\noindent
As we will show in our use case discussions, DBMSs often have opportunities to issue I/O requests in batches, for example, during buffer-pool eviction or group commits~\cite{DBLP:journals/pacmmod/LeisA0L023,DBLP:journals/pvldb/NarasayyaMSLSC15,DBLP:journals/pacmmod/NguyenAZL25}.
However, excessive batching can also introduce drawbacks; therefore, it must be tuned carefully to yield clear benefits.

\subsection{Inner Workings of \iouring{}}
\label{sec:innerworkings}
To effectively tailor DBMSs to \iouring{}, it is important to understand how I/O requests are executed.  
In the following, we examine the internals of \iouring{} and discuss other important aspects.

\paragraph{Issuing I/O requests.}
\iouring{} supports two ways to issue I/O in the kernel (step~(1) in \Cref{fig:iouring_work}).
In the default mode, the application thread (top left) calls \texttt{io\_uring\_enter} and transitions into kernel mode, where it processes submissions and completions.
The syscall can either block until a specified number of completions are available or return immediately.
In contrast, when applications set up \iouring{} with the \emph{SQPoll} mode, they avoid user--kernel transitions (syscalls) on the submission path by decoupling submission from execution.
A dedicated kernel thread (cf.\ \Cref{fig:iouring_work}, bottom left) continuously polls the SQ, issues I/O on behalf of the application, and posts results to the CQ.
When no new requests are submitted, the SQPoll thread enters a sleep state after a configurable timeout.
In a dedicated microbenchmark, we measured that waking this thread introduces a non-trivial latency of roughly 30~microseconds.
As our study later demonstrates, choosing the most beneficial execution strategy requires a thorough understanding of the application architecture and workload characteristics.

\paragraph{Execution paths in \iouring{}.}
I/O in \iouring{} can follow three main paths as shown in \Cref{fig:iouring_work} (steps 2a-c):

\noindent
\emph{(2a) Inline execution.}
When processing submissions, \iouring{} first attempts to complete requests inline, 
for example, when reading from a socket that has data already available.
Such operations execute immediately and their completion is posted to the CQ.

\noindent
\emph{(2b) Non-blocking execution.}
If an operation cannot be completed inline, its handling depends on its type.
For pollable operations, such as non-blocking socket reads, \iouring{} installs an internal event handler (\ttsafe{io_async_wake()}) that is executed when the socket becomes readable (\Cref{fig:iouring_work}, 2b).
By default, \iouring{} waits indefinitely for the operation unless a timeout via \ttsafe{OP_LINK_TIMEOUT} is set.

\noindent
\emph{(2c) Blocking execution.}
\label{sec:blocking_ops}
Certain operations cannot be executed asynchronously; for instance, blocking filesystem calls, such as \ttsafe{fsync}, or large storage reads.
In such cases, \iouring{} delegates execution to worker threads (\ttsafe{io_worker}).
This fallback is slower and incurs higher overhead than the native asynchronous paths.
Applications can explicitly request this behavior using the \texttt{IOSQE\_ASYNC} flag, which forces execution in a worker thread.
In a dedicated microbenchmark, issuing NOPs that were handled by \ttsafe{io_worker} threads added an average overhead of 7.3~microseconds compared to inline execution.
This additional cost results from offloading to a separate thread and synchronization between the worker and submission context.
Frequent fallback or a large number of active \texttt{io\_worker} threads typically indicates suboptimal I/O patterns and may warrant application-level redesign~\cite{axboe_liburing_issue1175_comment, axboe_liburing_issue189_comment, axboe_liburing_issue349_comment}.

\paragraph{Avoiding preemptions for completion-handling.}
When an asynchronous operation finishes, \iouring{} must run \emph{task\_work} in the kernel to place the completion entry into the CQ (step 3 in \Cref{fig:iouring_work}).
By default, this \emph{task\_work} runs whenever the application transitions from user to kernel space.
If the thread is busy (for example, during a join or scan), the kernel may issue an inter-processor interrupt (IPI) to process pending completions.
This effectively preempts the application, disrupts cache locality, increases jitter, and reduces batching efficiency.
To mitigate these effects, \iouring{} offers the \ttsafe{COOP_TASKRUN} flag (CoopTR), which reduces IPIs and allows applications to delay \emph{task\_work}.
However, completions are still processed on any kernel-user transition, including unrelated syscalls such as \hl{\ttsafe{mmap()}}.
Because preemptions have side effects, both the default and cooperative modes are ill-suited for modern high-performance DBMSs. 
\hl{
The \ttsafe{DEFER_TASKRUN} flag (DeferTR) runs \emph{task\_work} only on \texttt{io\_uring\_enter} calls, making it the recommended mode~\cite{iouringringperthread} because it gives applications more control and eliminates unwanted preemptions.
We therefore use it for the remainder of the paper, unless otherwise stated.
}



\paragraph{Other features for modern hardware.}
\iouring{} is designed to exploit modern hardware capabilities, supporting a variety of additional features.
These span high-level application optimizations and low-level runtime tuning for asynchronous I/O.
Key features include buffer registration and pinning to reduce memory management overhead, multishot operations, polling modes, and advanced request scheduling.
In the remainder of this paper, we examine how DB engines leverage these mechanisms to implement efficient I/O.

\section{Efficient Storage I/O with \iouring{}}
\label{sec:storage-intensive}

Modern NVMe devices can sustain millions of IOPS, yet conventional I/O stacks rarely reach this potential~\cite{DBLP:conf/damon/JasnyE0B25,DBLP:journals/pvldb/HaasL23}.
To explore if and how \iouring{} can close this gap, we discuss our three research questions (when to use, how to integrate, and how to tune \iouring{}) in the context of a buffer-managed storage engine. 

We follow a stepwise approach during its design to highlight how \iouring{}'s key capabilities impact system performance. 
After our use case discussion, we use targeted microbenchmarks to isolate specific \iouring{} behaviors in the storage context and provide insights for estimating achievable I/O gains.

\subsection{Use Case: Buffer-Managed Storage Engine}
\label{subsec:buffer-manager}

\begin{figure}
    \centering
    \captionsetup{aboveskip=0.0ex,belowskip=0.0ex}
    \captionsetup[subfigure]{aboveskip=0.0ex,belowskip=0.0ex}
    \includegraphics[width=0.75\linewidth]{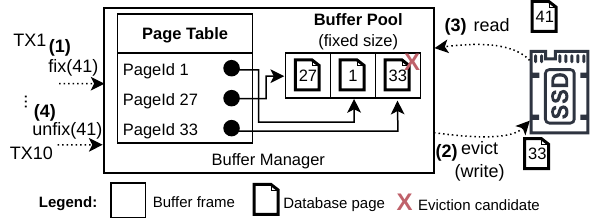}
    \caption{Overview of the buffer-managed storage engine design. Cold pages are evicted and written to disk freeing space that is used to cache frequently accessed pages.}
    \label{fig:bm_design}
\end{figure}

\hl{We use a buffer-managed storage engine as our primary use case because the buffer manager sits directly on the I/O path of out-of-memory transactions and orchestrates data movement between memory and SSDs.
We outline its core responsibilities and architecture (\Cref{fig:bm_design}), and then discuss how to exploit \iouring{} effectively.}

\paragraph{Buffer manager overview.}
A buffer manager caches frequently accessed pages and loads or evicts them as needed.
When a requested page is not present in the buffer pool, a \emph{page fault} triggers a read I/O to retrieve it from storage.
If the buffer pool is full, the buffer manager must select a page for eviction; if it is dirty, it is written back before its buffer frame is reused.
Because these operations lie on the critical transaction path, their efficiency is crucial for sustaining high throughput.
Although background tasks, such as checkpointing, also interact with the buffer manager and issue additional I/O, we ignore them for simplicity.


\paragraph{Buffer manager architecture.}
The buffer manager maintains a preallocated pool of buffer frames and a page table mapping logical page identifiers to frames, along with metadata such as reference bits and dirty flags.
It exposes two primitives~\cite{DBLP:journals/tods/EffelsbergH84}: 
\ttsafe{fix(page_id)} (1) checks whether the requested page resides in the buffer pool and loads it from storage (3) if not, evicting (2) another page if necessary, while \ttsafe{unfix(page_id)} (4) releases the page and marks it dirty if modified.
When the pool is full, a replacement policy selects victims. In this paper, we use clock-sweep \cite{PostgresClockSweep}, a common algorithm in DBMS:
pages are marked during the first pass and, if still unreferenced on the second pass, dirty pages are written back before the frame is reused.
The storage engine includes a B-tree index for tuple access and updates.
When the working set fits in memory, these operations complete without I/O; otherwise, page faults and evictions place I/O on the critical path, coupling buffer management with application logic.

\subsection{Workload \& System Conditions}
As mentioned in \Cref{sec:introduction}, the question \emph{when to use \iouring{}} -- when \iouring{} provides measurable performance gains -- depends on workload and system conditions.
\hl{We therefore use simple back-of-the-envelope models based on I/O cost and CPU utilization to estimate expected gains and validate the resulting system behavior.}

\begin{highlight}
\paragraph{Experimental setup.}
\phantomsection
\label{sec:exp-setup}
Unless stated otherwise, our experiments run on single-socket servers with an AMD EPYC 9654P processor (96 cores, SMT disabled), 768~GiB of DRAM, and Linux kernel 6.15.0 on Ubuntu 24.04 LTS with IOMMU disabled.
For storage experiments, one server is equipped with eight PCIe~5.0 NVMe SSDs (Kioxia CM7-R).
For network experiments, each node is equipped with a PCIe~5.0 Nvidia ConnectX-7 NIC (400~Gbit/s Ethernet) and connected through a 400~Gbit switch.
\end{highlight}

\paragraph{Workload conditions.}
To capture the impact of different workload characteristics, we use two standard DBMS benchmarks: single-statement, I/O-intensive YCSB-like transactions and the more complex, compute-bound TPC-C workload.
The experiments use a small 1~GB buffer pool.
For YCSB, we load 10~million tuples (8-byte key, 128-byte value), which with index structures and metadata results in a roughly 70\% page fault probability under uniform updates with 4~KiB pages, producing an I/O-bound workload well suited for storage analysis.
For TPC-C, we use 1 and 100 warehouses to study the effects of a mostly in-memory vs. a mostly out-of-memory setting.

\paragraph{System conditions.}
We evaluate several buffer manager configurations on one of these servers. 
The configurations range from fully synchronous to batched and asynchronous variants to study how the workloads interact with system conditions.
All configurations utilize a single-threaded setup, in which one core handles transaction processing and I/O requests.
This setup isolates I/O behavior from concurrency effects, allowing more precise analysis of performance improvements.
Later, in \Cref{sec:network-intensive_systems,sec:postgres}, we extend the analysis to multithreaded configurations.
\begin{highlight}
Our evaluation compares \iouring{} against conventional kernel-based I/O interfaces used in production systems.
We do not include kernel-bypass frameworks such as SPDK as direct baselines, since they target a different design space and would require substantial changes to the storage stack.
\end{highlight}

\subsection{Using \iouring{} in the Storage Engine}

\begin{figure}[]
    \centering 
    \captionsetup{aboveskip=0.0ex,belowskip=0.0ex}
    \captionsetup[subfigure]{aboveskip=0.0ex,belowskip=0.0ex}
    \includegraphics[width=0.95\linewidth]{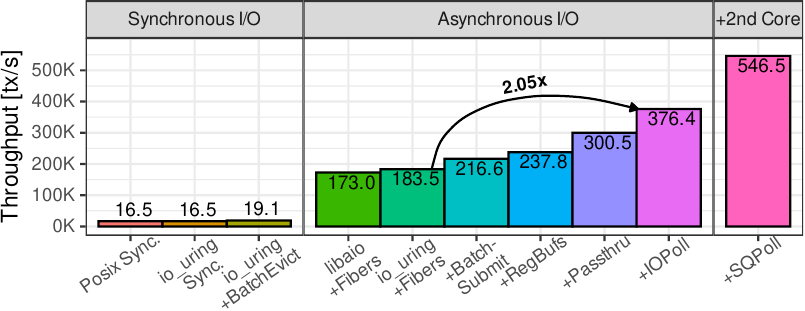}
    \caption{
    YCSB throughput (100\% uniform updates, one update per transaction) under different buffer manager designs and I/O execution modes.
    \iouring{} features and design optimizations are enabled incrementally from left to right, increasing transaction throughput from 16.5\,k to 546.5\,k~TPS.
    }
    \label{fig:bm_ycsb}
\end{figure}


Traditional buffer managers perform I/O through blocking system calls such as \texttt{pread()} and \texttt{pwrite()}, where each I/O request must complete before the thread can proceed.
As a performance baseline, we implement a synchronous buffer manager on top of \iouring{} by submitting one request at a time and waiting for its completion, ensuring that the DBMS thread has at most one outstanding I/O.
Although this baseline uses \iouring{}'s submission and completion queues for consistency with later variants, it does not exploit its asynchronous or batching features.

\paragraph{When \iouring{} does not help.} 
This setup is the simplest form of I/O execution, where the transaction thread blocks on every page fault or for eviction.
Consequently, throughput is directly tied to device latency, and using \iouring{} does not result in performance gains.
For the update-heavy YCSB workload mentioned earlier, our posix-based and \iouring{}-based implementation reaches a single-threaded throughput of 16.5\,k\,tx/s  (\Cref{fig:bm_ycsb}, \emph{Posix} and \emph{io\_uring}).
Because in-memory updates are negligible compared to storage latency, both implementations are \emph{I/O-latency bound}.

\paragraph{Modeling the bottleneck.}
\phantomsection
\label{sec:bm-modeling}
To validate the performance results, we use a simple latency-based model derived from the operation costs in Table~\ref{tbl:iocost}.
\hl{The numbers in Table~\ref{tbl:iocost} are measured independently using targeted experiments: the transaction execution cost is obtained from an in-memory run, the I/O costs are measured separately for the corresponding read/write paths, and only the 70\% page fault rate follows from the configured YCSB workload.}
Assuming a 70\% page fault rate and an average read-plus-write latency of $70+12=82$\,$\mu$s for our SSD device, the expected throughput is
$\frac{1}{0.7 \times 82 \times 10^{-6}} \approx 17.4\,\text{k\,tx/s}$.
The estimate aligns with the measured 16.5\,k\,tx/s, confirming that I/O latency rather than CPU or software overhead limits synchronous performance.
As a consequence, higher throughput can be achieved by reducing effective I/O latency or amortizing the latency of one request through batching.

\begin{table}[H]
\vspace{-0.5em}
\caption{I/O numbers used for performance modeling.}
\label{tbl:iocost}
\vspace{-1em}
\centering
\small
\setlength{\tabcolsep}{3pt}

\begin{minipage}[t]{0.31\linewidth}
\centering
\begin{tabular}{c|c}
    \shortstack{\textbf{Single}\\[-1.5pt]\textbf{Read}} & \shortstack{\textbf{Single}\\[-1.5pt]\textbf{Write}} \\
\hline
70\,$\mu$s & 12\,$\mu$s \\ 
\multicolumn{2}{c}{\textbf{\textsc{I/O Latency}}} \\
\end{tabular}
\end{minipage}
\hfill
\begin{minipage}[t]{0.68\linewidth}
\centering
\begin{tabular}{c|c|c|c}
    \shortstack{\textbf{Transaction}\\\textbf{Execution}} & \shortstack{\textbf{Single}\\[-1.5pt]\textbf{Read}} & \shortstack{\textbf{Batch}\\\textbf{Read}} & \shortstack{\textbf{Batch}\\\textbf{Write}} \\
\hline
8264\,clk & 10200\,clk & 5400\,clk & 5700\,clk \\ 
\multicolumn{4}{c}{\textbf{\textsc{CPU Cycles}}} \\
\end{tabular}
\end{minipage}
\vspace{-1em}
\end{table}

\subsubsection{Using \iouring{} to Batch Writes}
\label{subsec:batching_writes}
\hfill

\noindent
Synchronous I/O is bound by I/O latency, since each page fault triggers a blocking writeback of a dirty page before the next read can proceed.
Although \iouring{} cannot reduce device latency, its \emph{batching feature} can reduce the impact on the critical path.

\paragraph{Amortizing eviction cost.}
We leverage \iouring{} to introduce \emph{batched write submission} for the buffer manager's eviction path.
Instead of evicting and writing one page at a time, the buffer manager collects multiple victims and issues their writes together with a single \ttsafe{io_uring_enter()} call.
While execution remains synchronous and reads and writes do not yet overlap, batching lowers submission overhead and exploits device-level parallelism, demonstrating how a minor architectural change can benefit from \iouring{}'s strengths.

\paragraph{Performance implications.}
Batching write operations improves performance by about 14\%, reaching 19\,k\,tx/s (compare \Cref{fig:bm_ycsb}, \emph{+BatchEvict}) because submission overhead for eviction is amortized.
Eviction writes are issued in batches, so their latency is incurred once per batch rather than per eviction.
This amortizes the cost across $N$ evictions and leaves the 70$\mu$s read latency as the dominant term in the latency model from the previous section.
The expected throughput is
$\frac{1}{0.7 \times 70 \times 10^{-6}} \approx 20.4\,\text{k\,tx/s}$.
The measured and predicted results align closely, confirming that batching removes the write latency from the latency-bound path and effectively mitigates latency through amortization rather than elimination.


\subsubsection{Using \iouring{} for Asynchronous I/O} 
\label{subsec:async}
\hfill

\noindent
While batched writes amortize write latency, the buffer manager still operates synchronously, blocking on page faults and leaving the CPU idle during I/O.
Multiple threads, as studied in \Cref{subsec:storage_benchmarks}, hide this latency but introduce synchronization and scheduling overhead.
To hide latency in our single-threaded design, we therefore adopt \emph{asynchronous transaction execution} to overlap I/O and computation.

\paragraph{Overlapping compute \& I/O.}
\iouring{}'s completion-based model integrates naturally with async runtimes such as coroutines or fibers.
We extend the buffer manager with \texttt{fibers}~\cite{boostfiber} for cooperative scheduling.
Each transaction runs as a fiber that issues I/O requests and yields on page faults, allowing the \iouring{}-based runtime to schedule other fibers and keep the CPU active.

\paragraph{Cooperative transaction execution.}
Fiber context switches cost only tens of CPU cycles since they save and restore only register state, providing efficient user-level concurrency suited for I/O-intensive workloads.
When a suspended fiber's I/O completes, it is marked ready and resumed by the scheduler.
Since all concurrency is cooperative, the B-tree implementation requires neither locks nor atomic operations across fibers.
If a fiber resumes after an I/O delay and the data structure has changed, it restarts the B-tree traversal to ensure correctness and preserve isolation without explicit synchronization. Considering such details is important for accurate performance modeling, as shown later.

\paragraph{CPU, the new bottleneck.}
With up to 128 fibers, throughput rises by nearly an order of magnitude to 183\,k~tx/s (\Cref{fig:bm_ycsb}, \emph{+Fibers}).
Under the same asynchronous execution scheme, \emph{libaio} reaches 173\,k~tx/s, so we continue the analysis on \iouring{}, which provides higher throughput and exposes additional optimization opportunities.
At this concurrency level, the system becomes CPU- rather than latency-bound: concurrent fiber execution hides I/O latency and the CPU is fully utilized.
We therefore switch from a latency- to a cycle-based model that accounts for per-transaction CPU cost.

\hl{Using hardware cycle counters (\texttt{rdtsc}), we measure transaction logic (B-tree traversal and update) as $c_\mathrm{tx}=8{,}264$ cycles, and I/O processing (submission and completion) as $c_\mathrm{io}=c_{\text{read-single}}+c_{\text{write-batch}}=15{,}900$ cycles (\Cref{tbl:iocost}).}
For a 3.7\,GHz core and a page fault rate of $r_{pf}=70\%$, the expected throughput is
$\frac{\text{clock frequency}}{c_{tx} + r_{pf} \times c_{io}} = 
\frac{3.7\times10^9}{8264 + 0.7\times15900} \approx 190.8\,\text{k\,tx/s},$ 
matching the measured 183\,k\,tx/s.
This confirms that CPU overhead, rather than I/O latency, now dominates performance, as intended with asynchronous execution.

\subsubsection{Using \iouring{} to Batch Reads}
\label{subsec:batching_reads}
\hfill

\noindent
In the initial asynchronous design, each fiber submits its I/O request right before blocking and is woken up once the I/O completes.
This hides I/O latency by overlapping reads and writes but incurs syscall overhead for each I/O.
We therefore introduce \emph{batched read submission}, which groups read requests from multiple fibers before entering the kernel via \ttsafe{io_uring_enter()}.
The batched submission amortizes syscall overhead and exploits device-level parallelism, reducing per-I/O cycle cost (\Cref{tbl:iocost}) and improving CPU efficiency.

\paragraph{Adaptive batching.}
Read batching improves throughput by lowering per-I/O cost but may introduce queuing delays if the runtime waits too long to collect requests, while very small batches negate the amortization benefit.
Our runtime therefore uses \emph{adaptive batching}, adjusting the batch size based on the ratio of outstanding I/Os to waiting fibers.
When many I/Os are in flight, additional submissions are deferred to increase amortization; when few are pending, batches are flushed earlier to keep the CPU busy.
\hl{This feedback mechanism maintains high device utilization while avoiding stalls when the scheduler runs out of runnable fibers.}

\paragraph{Impact of adaptive batching.}
Performance increases by about 18\%, from 183\,k to 216\,k\,tx/s (\Cref{fig:bm_ycsb}, \emph{+BatchSubmit}).
Adding I/O cost for batched reads to the model,
$c_\mathrm{io}=c_{\text{read-batch}}+c_{\text{write-batch}}=11{,}100$ cycles (\Cref{tbl:iocost}),
yields $\frac{3.7\times10^9}{8264 + 0.7\times11100} \approx 230\,\text{k\,tx/s}$ as expected throughput.
The estimate aligns with the measured 216\,k\,tx/s, confirming that adaptive read batching reduces CPU overhead in the submission path and improves single-core efficiency.

\begin{figure}[]
  \centering
  \captionsetup{aboveskip=0.0ex,belowskip=0.0ex}
  \captionsetup[subfigure]{aboveskip=0.0ex,belowskip=0.0ex}
    \includegraphics[width=0.95\linewidth]{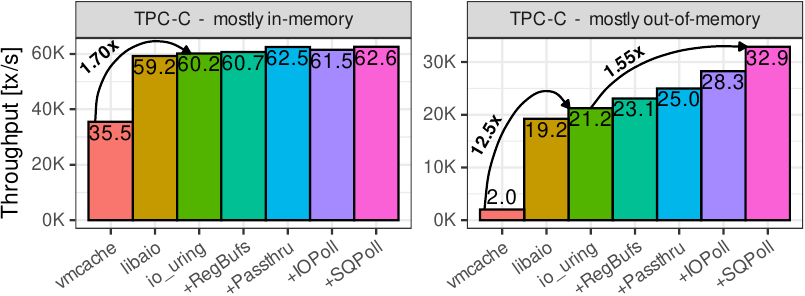}
    \caption{TPC-C with 1 warehouse (left) and 100 warehouses (right) with the default transaction mix. \iouring{} outperforms libaio, vmcache uses blocking I/O that performs worst if out-of-memory with 100 warehouses (storage-intensive).} 
  \label{fig:bm_tpcc}
\end{figure}

\subsection{Tuning \iouring{} for the Storage Engine}

\noindent
Utilizing \iouring{}'s key capabilities enabled us to implement an asynchronous architecture whose performance is determined by CPU cycles spent on I/O processing.
This forms the basis for applying low-level features to reduce I/O overheads.
However, the effectiveness of \iouring{}'s features also depends on workload characteristics, as discussed next. 
\hl{The applicability of these features to multi-threaded execution is evaluated separately in \Cref{subsubsec:bm_multithreaded}}.

\pagebreak 
\subsubsection{Performance Evaluation with YCSB}
\label{subsubsec:io_optimizations_yscb}
\hfill

\noindent
We cut per-I/O CPU cost by reducing three overheads: data movement (using registered buffers), storage stack (NVMe passthrough), and submission/completion handling (SQPoll \& IOPoll).

\paragraph{Registered buffers reduce copies.}
\phantomsection
\label{sec:storage_reg_buffers}
\iouring{} allows user-space buffers to be \emph{registered} once during initialization, \hl{avoiding pinning, DMA setup, and kernel-user copies on every I/O.}
The kernel then performs DMA directly into user memory, eliminating these overheads.
For YCSB, this zero-copy optimization improves throughput by about 11\%, reaching 238\,k\,tx/s (\Cref{fig:bm_ycsb}, \emph{+RegBufs}). 

\paragraph{NVMe passthrough skips abstractions.}
\phantomsection
\label{sec:nvme_passthru}
To access NVMe devices directly, \iouring{} provides the \ttsafe{OP_URING_CMD} opcode, which issues native NVMe commands via the kernel to device queues. 
\hl{Compared to kernel-bypass frameworks such as SPDK, these optimizations can be adopted incrementally on top of the regular kernel I/O stack.}
By bypassing the generic storage stack, passthrough reduces software-layer overhead and per-I/O CPU cost.
This yields an additional 20\% gain, increasing throughput to 300\,k\,tx/s (\Cref{fig:bm_ycsb}, \emph{+Passthru}).

\paragraph{IOPoll avoids interrupts.}
With \texttt{IOPOLL}, completion events are polled directly from the NVMe device queue, either by the application or by the kernel \texttt{SQPOLL} thread (cf. \Cref{sec:background}), replacing interrupt-based signaling.
This removes interrupt setup and handling overhead but disables non-polled I/O, such as sockets, within the same ring.
When using filesystems, \texttt{IOPOLL} requires explicit support and is typically available only for direct block-device access via \ttsafe{O_DIRECT}.
As shown in \Cref{fig:bm_ycsb} (\emph{+IOPoll}, right), completion polling provides an additional 21\% throughput gain, reaching 376\,k\,tx/s - single-threaded.
As we will show later, it also reduces latency for I/O-intensive workloads (cf. \Cref{fig:fsync_lat}).

\paragraph{SQPoll eliminates syscalls.}
In \texttt{SQPOLL} mode, a dedicated kernel thread continuously polls the submission queue, allowing applications to enqueue I/O requests without calling \ttsafe{io_uring_enter()} for each submission.
This dedicates one CPU core to polling but eliminates most syscall and submission overheads. 
The kernel thread handles I/O completions and places them into the completion queue for later consumption by the application.
For our buffer manager (\Cref{fig:bm_ycsb}, \emph{+SQPoll}), throughput increases by about 32\% to 546k~tx/s, corresponding to the cost previously spent in syscall and kernel-side processing.

\subsubsection{Performance Evaluation with TPC-C}
\label{subsubsec:io_optimizations_tpcc}
\hfill


\noindent
Unlike YCSB, which issues short, independent transactions dominated by random I/O, TPC-C models an OLTP system with interacting transactions and a larger share of in-memory computation.
This workload thus evaluates how \iouring{} optimizations perform in a less I/O-bound, more CPU-intensive setting.
As a baseline, we use \texttt{vmcache}~\cite{DBLP:journals/pacmmod/LeisA0L023}, a state-of-the-art buffer manager.
We reuse the asynchronous, batched-read configuration from the YCSB experiments for a fair comparison.

\paragraph{Workload-dependent benefits.}
\Cref{fig:bm_tpcc} shows the results for TPC-C in a mostly in-memory and a mostly out-of-memory configuration.
The \iouring{}-based buffer manager consistently outperforms \texttt{vmcache}, achieving up to 12.5$\times$ higher throughput in the naive configuration.
The primary reason is architectural: \texttt{vmcache} relies on blocking reads.
Enabling advanced \iouring{} features further improves throughput, although the relative gains depend on the workload configuration.
The memory-intensive TPC-C workload is largely compute-bound and many reads can be answered from memory which limits the effect of I/O-path optimizations.
In this case, \emph{+IOPoll} performs slightly worse than the interrupt-driven baseline because polling wastes CPU cycles when I/O operations are sporadic.
\emph{+SQPoll} provides no benefit for the same reason.
In the out-of-memory setting, I/O activity rises due to the increased cost of page loads and evictions, making optimizations more impactful.

\begin{highlight}
    
\subsubsection{Multi-Threaded Performance Evaluation}
\label{subsubsec:bm_multithreaded}
\hfill

\begin{figure}[]
  \centering
  \captionsetup{aboveskip=0.0ex,belowskip=0.0ex}
  \captionsetup[subfigure]{aboveskip=0.0ex,belowskip=0.0ex}
    \includegraphics[width=0.95\linewidth]{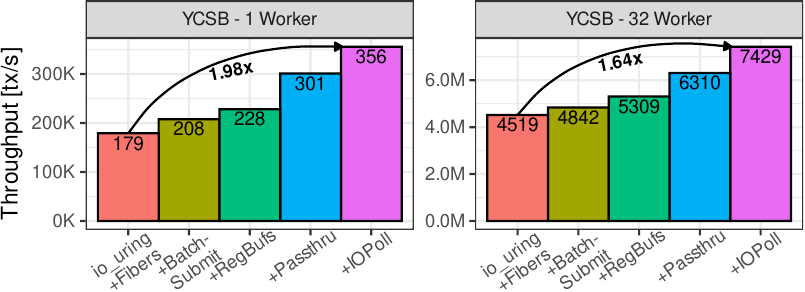}
    \caption{\iouring{} in a multi-threaded buffer manager. Single worker numbers (left) as reference for additional concurrency logic. Optimizations have a similar impact on runtime performance when multiple worker threads are used (right). } 
  \label{fig:bm_ycsb_multithreaded}
\end{figure}

\noindent
To study whether the \iouring{} optimizations carry over to parallel execution, we extend the buffer manager with a multi-threaded runtime.
The runtime uses the same I/O paths as earlier and only introduces additional contention and synchronization handling for worker threads.
We use an optimistic-lock-coupling B-tree together with an atomic page table for low-overhead concurrency control.
To further reduce synchronization overhead, eviction is organized per shard, and each worker has a dedicated eviction fiber using the same clock-sweep algorithm as in the single-threaded design.
Such optimizations are critical to prevent I/O synchronization bottlenecks from masking end-to-end I/O performance impacts.


\paragraph{Relative benefits remain the same.}
In \Cref{fig:bm_ycsb_multithreaded}, we report the results of the previous YSCB benchmark when executed using a single worker (performance after multi-threading was added) and 32 worker threads.
The results confirm that due to the effective contention and synchronization handling in our multi-threaded buffer manager above the I/O layer, the relative effect of the \iouring{} optimizations remains similar in the multi-threaded setting.

\end{highlight}

\subsection{Take-aways and Summary}

\paragraph{When to use \iouring{}.}
Our buffer manager study shows that \iouring{} yields meaningful gains for I/O-intensive workloads with many page faults, such as YCSB and TPC-C configurations where a substantial fraction of reads miss the buffer pool and involve SSD page loads and evictions.
In compute-heavy (i.e., mostly in-memory) settings, the I/O path contributes little to overall cost, and \iouring{} optimizations have correspondingly smaller impact.

\paragraph{How to integrate it.}
A key insight of our study is that \iouring{} must be integrated as part of an end-to-end architectural design rather than as a drop-in replacement.
Using \iouring{} key features (batching, asynchronous execution) enabled us to shift the bottleneck from device latency to CPU cycles and make the cost of I/O processing explicit, as verified by our model-based analysis.

\paragraph{How to tune it.}
Once the architecture exposes enough asynchronous I/O, low-level \iouring{} features can reduce per-operation CPU overhead.
However, our study revealed that the tuning benefits depend strongly on the workload: substantial improvements arise when I/O dominates execution time, while CPU-bound or cache-resident workloads gain little.

%

\subsection{Detailed Analysis of \iouring{}}
\label{subsec:storage_benchmarks}

The results in the previous section showed that realizing performance gains depends on both the system architecture and workload characteristics. 
In this section, we conduct targeted microbenchmarks to isolate and quantify the effects of individual features for system builders in depth and also study effects of multi-threading.

\begin{figure}[]
    \centering
    \captionsetup{aboveskip=0.0ex,belowskip=0.0ex}
    \captionsetup[subfigure]{aboveskip=0.0ex,belowskip=0.0ex}
    \includegraphics[width=0.99\linewidth]{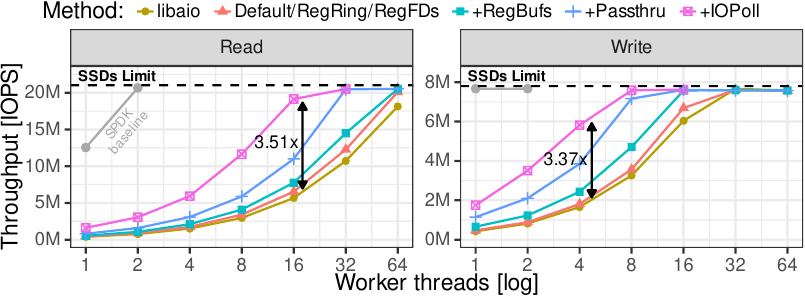}
    \caption{\hl{Scale-up performance for random 4~KiB reads \& writes between \ttsafe{libaio} and \iouring{} with incremental optimizations. \iouring{} consistently outperforms \ttsafe{libaio}. Further \iouring{} optimizations increase throughput by 3.4--3.5$\times$.}}
    \label{fig:ssd_scaleout}
\end{figure}

\paragraph{Batching effects on SSD latency.}
In the buffer manager design, we used \iouring{}'s batching to hide I/O latency for writes and amortize syscall overhead for reads.
However, batching can also increase latency variance, which is problematic for workloads requiring predictable response times.
To quantify this effect, we vary batch sizes for write requests to a single SSD via \iouring{}, fixing throughput at 1.5~MIOPS to stay below the device limit and isolate batching behavior.
As shown in \Cref{tbl:ssdbatchlat}, small batches (e.g., size~8) keep latencies mostly below 25~µs, at the cost of slightly higher syscall frequency.
Larger batches reduce submission overhead but cause higher variance; with batch size~128, latency spikes up to 200~µs occur as bursts from multiple workers can temporarily overload the SSD queue with many outstanding I/Os.
Thus, even below I/O saturation, batch size strongly influences latency distribution.
For latency-sensitive DBMSs, overly aggressive batching is counterproductive. 

\begin{table}[H]
\vspace{-0.5em}
\caption{Impact of batch size on SSD write latency (8 workers).}
\label{tbl:ssdbatchlat}
\vspace{-1em}
\small
\setlength{\tabcolsep}{4pt}
\centering
\begin{tabular}{l|cccccc}
Batch size: & 1 & 8 & 32 & 64 & 128 & 256 \\
\hline
Latency $\oslash$ [µs] & 11.51 & 24.22 & 60.62 & 116.40 & 200.85 & 317.51 \\
Latency $\sigma$ [µs] & ±0.95 & ±1.71 & ±3.91 & ±12.17 & ±7.47 & ±33.88 \\
\end{tabular}
\vspace{-1em}
\end{table}

\paragraph{Multi-threaded performance.}
\hl{To analyze how \iouring{} behaves under parallelism in isolation, we now increase the number of I/O threads.}
We use one ring-per-thread for \iouring{} and also include \texttt{libaio} for comparison.
\hl{This experiment isolates the scalability of the I/O path and does not use the buffer manager framework from the previous section.}
As shown in \Cref{fig:ssd_scaleout}, \iouring{} consistently outperforms \texttt{libaio} for both random reads and writes and exhibits near-linear scalability with the number of threads. 
At higher core counts, the operating system's storage stack becomes the dominant bottleneck.
The benefit of \iouring{} optimizations increases with scale: registered buffers (+RegBufs) reduce CPU overhead, while NVMe passthrough (+Passthru) and IOPoll in particular deliver substantial throughput improvements of 3.4--3.5$\times$ saturating the SSD array with 18 and 6 cores, respectively. 

\begin{figure}[]
  \centering
  \captionsetup{aboveskip=0.0ex,belowskip=0.0ex}
  \captionsetup[subfigure]{aboveskip=0.0ex,belowskip=0.0ex}
    \includegraphics[width=0.99\linewidth]{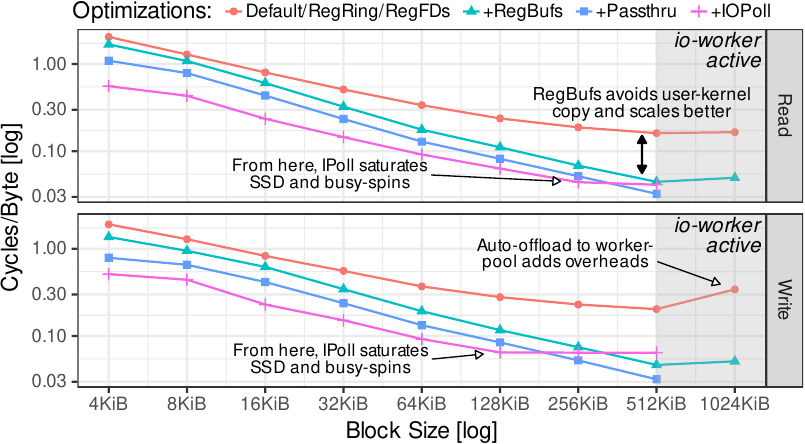}
    \caption{Single thread SSD performance with increasing block sizes. Performance degrades with I/O workers, NVMe-passthrough is only supported until 512KiB.} 
  \label{fig:ssd_read_bs}
\end{figure}

\paragraph{Increasing block sizes.}
Larger block sizes for I/O can further amortize CPU costs when the workload allows coarser-grained access.
We therefore evaluate how block size affects SSD performance by measuring CPU cycles per byte for reads and writes while varying the I/O block size in a single-threaded setup.  
\Cref{fig:ssd_read_bs} shows that larger blocks substantially reduce CPU cost per byte, as syscall and I/O stack overheads are amortized.  
With sufficiently large requests, a single core saturates the PCIe~5 SSD array, reaching up to 90~GiB/s for reads and 50~GiB/s for writes, close to hardware limits.
This point is reached for writes at 128~KiB and reads at 256~KiB with +Passthru and +IOPoll enabled.

\paragraph{Large blocks can backfire.}
However, exceeding certain thresholds triggers asynchronous worker threads, signaling fallback to slower I/O paths as discussed in \Cref{sec:innerworkings}.
First, if the block size exceeds \ttsafe{max_hw_sectors_kb} (which can be 128~KiB if the IOMMU were enabled), workers are spawned even at low I/O depth, as a single request surpasses the maximum DMA size.
Second, with \ttsafe{O_DIRECT}, workers appear once the number of batched requests exceeds \ttsafe{nr_requests} (1023 on bare metal, 127 in our cloud VM), and on some consumer SSDs even without \ttsafe{O_DIRECT}.
Third, when block sizes exceed 512~KiB (\ttsafe{max_segments}) asynchronous workers are again used internally for I/O.
While large blocks improve efficiency and fully utilize PCIe~5 bandwidth, surpassing these software or hardware limits causes worker fallback, reintroducing latency and CPU overhead in \iouring{}.

\begin{figure}[]
  \centering
  \captionsetup{aboveskip=0.0ex,belowskip=0.0ex}
  \captionsetup[subfigure]{aboveskip=0.0ex,belowskip=0.0ex}
    \includegraphics[width=0.95\linewidth]{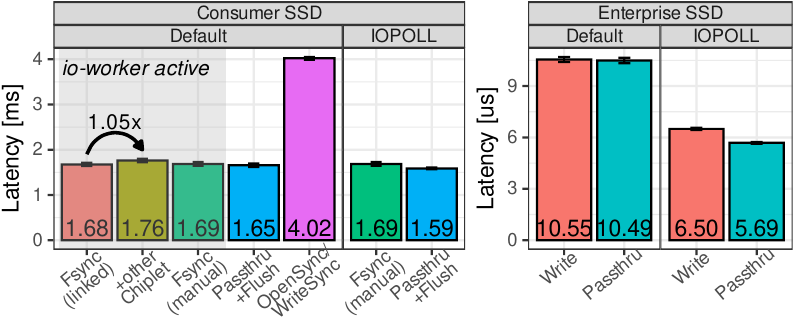}
    \caption{Durable writes with \iouring{}. Left: Writes and fsync are issued via \iouring{} or manually linked in the application. Right: Enterprise SSDs do not require explicit fsync.}
  \label{fig:fsync_lat}
\end{figure}

\paragraph{The durable write problem.}
Durable writes are essential for database systems, particularly for write-ahead logging and checkpointing, yet remain costly.
The standard approach, \emph{fsync}, is blocking in \iouring{} and thus executed by fallback worker threads.  
Moreover, \emph{fsync} cannot be issued by rings configured for IOPoll and must be used from a separate ring or as a traditional syscall.
These constraints motivate alternatives such as opening files with \ttsafe{O_SYNC}, which delegates durability to the kernel, or using NVMe flushes via passthrough commands.
\Cref{fig:fsync_lat} compares these methods on consumer and enterprise SSDs \hl{with \ttsafe{O_DIRECT}}.
Enterprise SSDs with DRAM caches and Power Loss Protection (PLP) achieve microsecond-level latencies, whereas consumer SSDs remain dominated by intrinsic millisecond-scale \emph{fsync} cost, masking worker-thread latencies.
Not pinning the \emph{fsync} I/O worker to the local chiplet (+Chiplet) increases latency by about 5\%.
\ttsafe{O_SYNC}-based writes perform poorly, more than twice as slow as explicit writes followed by \emph{fsync}.  
NVMe passthrough with explicit flush commands offers a truly asynchronous durability path but requires raw device access.  
Linking a write and \emph{fsync} in \iouring{} offers no improvement over issuing them sequentially in the application.  
For enterprise SSDs, durability is managed by the device itself, eliminating the need for \emph{fsync}.  
Passthrough writes with IOPoll reduce latency by about one microsecond, showing the benefit of bypassing the storage stack in latency-sensitive workloads.
While passthrough with flush is the most efficient asynchronous option, its lack of filesystem support restricts it to flash-optimized database systems.  
Implementing durable writes in \iouring{} therefore requires careful configuration to avoid performance pitfalls.

\section{Efficient Network I/O with \iouring{}}
\label{sec:network-intensive_systems}

After analyzing storage, we now focus on the networking aspects of \iouring{}.
High-speed interconnects with link rates in the range of 400 Gbit/s are now a commodity in modern data centers and cloud deployments~\cite{aws200G, DBLP:journals/pvldb/WuCCIS25}.
\hl{We study how distributed DBMSs can exploit \iouring{} in these high-throughput settings and complement the use-case analysis with targeted microbenchmarks.}

\subsection{Use Case: Distributed Data Shuffle}
\label{subsec:shuffle}

Shuffle operations are a key building block in any distributed DBMS.
We now study a bi-directional, all-to-all data shuffle representative of distributed analytical query execution, as used in parallel joins.

\paragraph{Distributed shuffle overview.}
\hl{In a distributed shuffle, each node scans local tables and repartitions tuples across nodes according to a partitioning function.
For a distributed hash join, tuples are shuffled by join key. Incoming tuples are inserted into the local probe table, and tuples of the larger relation are later probed there.}

\paragraph{Shuffle engine architecture.}
We implement a distributed hash join using morsel-based processing~\cite{DBLP:conf/sigmod/LeisBK014}.
\hl{Each worker repeatedly fetches morsels, partitions tuples, and issues send and receive operations while also performing the local hash-table build.}
%
%
\subsection{Workload \& System Conditions}
\hl{Like the buffer manager, the distributed shuffle combines computation and I/O that compete for CPU cycles, but its workload and system conditions differ substantially.}

\paragraph{Workload conditions.}
\hl{To study \emph{when} \iouring{} yields measurable benefits, we vary tuple size while keeping the total shuffle volume fixed.
Smaller tuples cause more probe-table inserts per transferred byte and therefore more random memory accesses, whereas larger tuples reduce insertion frequency and memory pressure.}

\begin{figure}
    \centering
    \captionsetup{aboveskip=0.0ex,belowskip=0.0ex}
    \captionsetup[subfigure]{aboveskip=0.0ex,belowskip=0.0ex}
    \includegraphics[width=0.85\linewidth]{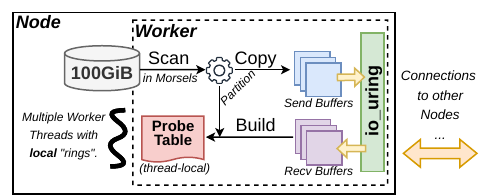}
    \caption{Overview of the shuffle architecture with scan \& probe table. Workers use morsel-driven parallelism and handle scanning, probe table building and network I/O.}
    \label{fig:shuffle_design}
\end{figure}

\paragraph{System conditions.}
\phantomsection
\label{sec:exp-setup-shuffle}
\hl{We use the previously described six-node cluster, providing 4.8~Tbit/s of bisection bandwidth (6$\times$2$\times$400~Gbit/s), to evaluate shuffle workloads.}
Since network round-trip latencies are orders of magnitude higher than SSD access latencies, even in traditional engines, it is common to adopt asynchronous strategies such as using dedicated I/O threads~\cite{DBLP:conf/nsdi/DragojevicNCH14,DBLP:conf/nsdi/KaliaKA19,mysqlaio}.
Batching tuples into larger chunks, e.g., 1~MiB, is also an often adopted approach to amortize network overhead and increase throughput~\cite{DBLP:journals/pvldb/JepsenLPSC21}.
Hence, we adopt an asynchronous design as a baseline without discussing synchronous I/O in the context of the shuffle use case.

\subsection{Using \iouring{} in the Shuffle Engine}
\label{sec:shuffle_design}
Building on these insights and the results of the buffer manager use case in \Cref{sec:storage-intensive}, we directly utilize \iouring{}'s asynchronous I/O and batching capabilities for a first shuffle baseline.

\paragraph{Integrating \iouring{} into a shuffle.}
\phantomsection
\label{sec:skew-thread-per-ring}
The asynchrony of \iouring{} enables us to avoid the approach of using dedicated I/O threads that use blocking I/O. 
As recommended with \iouring{}~\cite{iouringringperthread}, we adopt a ring-per-thread architecture in which each worker thread possesses a thread-local \iouring{} ring.
\hl{Skew is handled at a higher level in the application, not at the I/O submission level.}
This architecture enables scanning data, while sending and receiving tuples asynchronously via \iouring{} within a single thread.
While both approaches overlap computation and communication, increasing CPU utilization, co-locating computation and I/O within the same thread avoids synchronization overhead with I/O workers and improves cache locality.
Unlike in our buffer manager, we directly employ multithreading with the goal to saturate the 400~Gbit/s NICs.

\paragraph{Details of shuffle implementation.}
To avoid other system bottlenecks and isolate \iouring{} performance impacts, we adopt several state-of-the-art system optimizations in our shuffle implementation.
For example, we use batched inserts into the probe table to improve access predictability and allow the prefetcher to issue concurrent loads~\cite{DBLP:conf/damon/BirlerSF024}.
We pin workers with their TCP/IP flows in a round-robin manner to CPU \emph{chiplets}~\cite{DBLP:journals/pvldb/FogliZPBG24} to reduce overheads in \iouring{} for networking.
Finally, we tune our network stack following industry best practices~\cite{azurenetguidelines} and use Linux 6.17 for zero-copy recv support.

\paragraph{Basic \iouring{} configuration.}
\label{sec:shuffle_impl}
\iouring{} comes with multiple network-specific features, such as multi-shot receive (cf. \Cref{subsec:net_benchmarks}), that can be used for data shuffling.
As a baseline, we adopt the same DeferTR and single-issuer setup flags as used in the buffer manager use case to avoid preemptions and internal synchronization (cf. \Cref{sec:background}).
For transfers, we use large buffer sizes of 1~MiB, which are more suitable than multi-shot receive or other mechanisms. 

\begin{figure}[]
  \centering
  \captionsetup{aboveskip=0.0ex,belowskip=0.0ex}
  \captionsetup[subfigure]{aboveskip=0.0ex,belowskip=0.0ex}
  \includegraphics[width=0.99\linewidth]{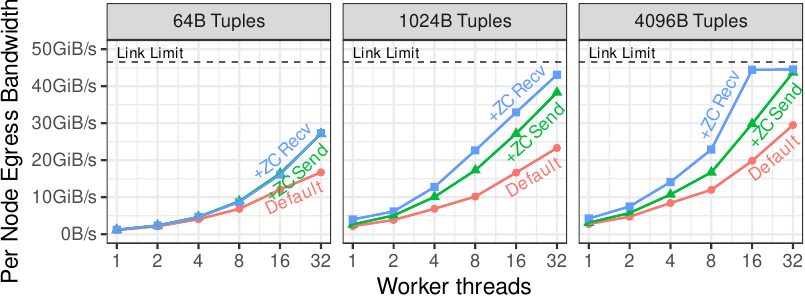}
  \caption{Per-node egress bandwidth for a six node shuffle with different tuple sizes using up to 32 worker threads.
  }
  \label{fig:shuffle_scaleout}
\end{figure}

\paragraph{Initial performance.}
\Cref{fig:shuffle_scaleout} shows the per-node egress bandwidth for small and large tuple sizes as the number of worker threads increases.
The total bidirectional bandwidth is therefore twice these values.
The baseline \iouring{} configuration (red) scales with more workers, mainly due to the morsel-based design, where workers can send and receive data independently.
\hl{For small tuple sizes (e.g., 64~B), frequent probe-table inserts introduce random memory accesses that stall the CPU and limit network throughput, whereas larger tuples reduce insertion pressure and achieve higher bandwidths.}
However, even for larger tuples, the system reaches at most 30~GiB/s per node (\Cref{fig:shuffle_scaleout}, red), well below the 400~Gbit/s link rate, indicating that the workload is not I/O-bound.

\subsection{Tuning \iouring{} for Network I/O}
As with storage, \iouring{} also provides multiple network optimizations to reduce per-I/O overhead.
In the following, we discuss our findings on how to best use them for efficient network shuffling.

\paragraph{Reducing memory pressure.}
\hl{At high network speeds, copies between user-space and kernel buffers become costly because even a single extra copy consumes substantial memory bandwidth.
To reduce this overhead, \iouring{} supports zero-copy send and receive, transmitting data directly from or into pinned user-space buffers.}

\begin{figure}[]
    \centering
    \captionsetup{aboveskip=0.0ex,belowskip=0.0ex}
    \captionsetup[subfigure]{aboveskip=0.0ex,belowskip=0.0ex}
    \includegraphics[width=0.99\linewidth]{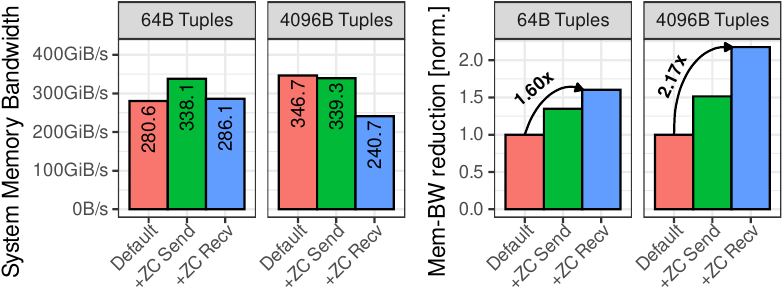}
    \caption{Memory bandwidth for the data shuffle in Fig.~\ref{fig:shuffle_scaleout} (32 Workers).
    Left: absolute bandwidth; right: Bandwidth reduction normalized by achieved network throughput.}
    \label{fig:shuffle_mem}
\end{figure}

\paragraph{Zero-copy I/O reduces memory load.}
As shown in \Cref{fig:shuffle_scaleout}, enabling zero-copy send (green) and zero-copy receive (blue) yields visible throughput improvements across different tuple sizes compared to the default setting.
For 4~KiB tuples, link bandwidth is saturated with only 16 workers per node when both zero-copy paths are active, achieving full bidirectional 400~Gbit/s utilization with modest CPU usage.
However, for 64B tuples, zero-copy receive does not provide additional benefits beyond zero-copy send.
We suspect that contention between NIC traffic and CPU-driven random memory accesses diminishes the potential gains, but a deeper analysis of zero-copy receive is left for future work.

\paragraph{Analyzing memory bandwidth.}
To understand why shuffle performance is limited without zero-copy operations, we analyze system memory bandwidth during the \hl{scale-up} experiment described above.
\Cref{fig:shuffle_mem} shows measurements for both the default and zero-copy configurations.
On the left, we present the system memory bandwidth, calculated as the sum of read and write traffic from hardware performance counters.
For both workloads, peak system bandwidth approaches 400~GiB/s.
Zero-copy configurations typically show equal or higher absolute memory bandwidth, primarily due to their higher network throughput, which increases overall memory traffic.
To account for this effect, we normalize the measured bandwidth by the achieved network throughput.
The normalized plot (cf. \Cref{fig:shuffle_mem}, right) shows that using zero-copy for send and receive reduces effective memory bandwidth by about half, as expected, since data copies are eliminated in both directions.

\begin{figure}[]
  \centering
  \captionsetup{aboveskip=0.0ex,belowskip=0.0ex}
  \captionsetup[subfigure]{aboveskip=0.0ex,belowskip=0.0ex}
  \includegraphics[width=0.99\linewidth]{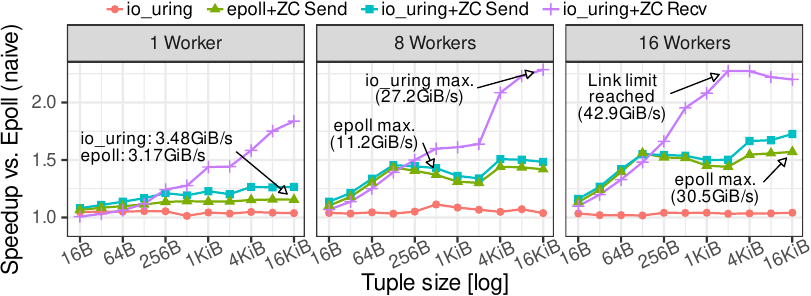}
  \caption{Speedup of \iouring{} and zero-copy send epoll vs. plain epoll for data shuffling across six nodes and different tuple sizes. Zero-copy receive is only available with \iouring{}.} 
  \label{fig:shuffle_epoll}
\end{figure}

\paragraph{End-to-end comparison with epoll.}
Similar to the buffer manager, we compare our optimized \iouring{}-enabled implementation to state-of-the-art approaches.
For the shuffle, we use epoll as a baseline, a readiness-based I/O interface commonly used in existing systems.
\Cref{fig:shuffle_epoll} shows the speedup of different \iouring{} variants and an epoll-based zero-copy send implementation against a naive epoll baseline without optimizations.
We vary tuple sizes (smaller tuples result in more random inserts into the probe table per byte transferred) and scale up to sixteen workers to avoid link saturation and ensure a fair comparison.
Without zero-copy, epoll is only marginally slower than \iouring{} despite issuing more system calls.
This behavior stems from both implementations transferring tuples in large 1 MiB chunks, which amortizes syscall and I/O-path overhead.
With zero-copy send, \iouring{} achieves substantially better performance, with the gap widening as the number of workers increases.
Unlike epoll, \iouring{} also supports zero-copy receive, further improving performance and providing up to a 2.5$\times$ speedup.

\begin{highlight}
\paragraph{Unified storage and network I/O.}
\phantomsection
\label{sec:unified-io}
To evaluate the practical benefit of \iouring{}'s unified asynchronous interface, we measure a remote table scan in which a server reads data pages from SSD and sends them to a client over the network.
With \iouring{}, storage and network operations are submitted and completed through the same ring, whereas \texttt{epoll}+\texttt{libaio} must coordinate separate storage and network subsystems and their event handling.
The results in \Cref{fig:remotescan} show that \iouring{} (green+blue) consistently outperforms the split \texttt{libaio}+\texttt{epoll} design (red), with the gap widening further once registered buffers and zero-copy are enabled, while for larger page sizes the difference narrows because costs are amortized.

\begin{figure}[]
  \centering
  \captionsetup{aboveskip=0.0ex,belowskip=0.0ex}
  \captionsetup[subfigure]{aboveskip=0.0ex,belowskip=0.0ex}
    \includegraphics[width=0.99\linewidth]{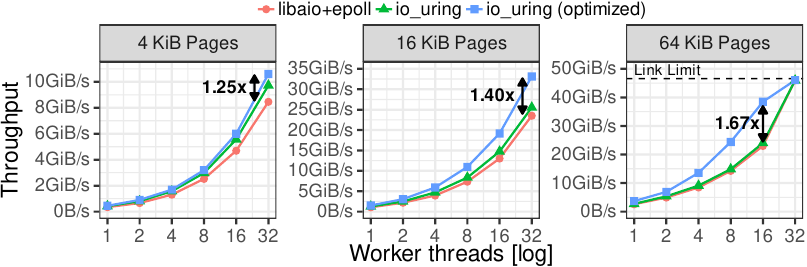}
    \caption{Remote table scan combining storage and network I/O. A client requests pages of different sizes over 32 concurrent connections per worker from a server that reads them from SSD and returns them over the network.}
    \label{fig:remotescan}
\end{figure}

\end{highlight}

\subsection{Take-aways and Summary}

\paragraph{When to use \iouring{}.}
Our shuffle use case shows that \iouring{} provides benefits in high-throughput settings where memory bandwidth becomes a bottleneck.
With large tuples, when probe table inserts are rare, and workers mainly stream tuples and perform simple partitioning, zero-copy networking enables us to saturate 400~Gbit/s links with relatively few cores.
For smaller tuple sizes, random memory accesses during hash-table inserts dominate and limit throughput; network-path optimizations have a less significant impact but remain important for achieving optimal bandwidth.

\paragraph{How to integrate it.}
Treating \iouring{} as a drop-in replacement in a traditional I/O-worker design is inadequate.
It requires a ring-per-thread design that overlaps computation and I/O within the same thread.
Combined with careful placement of workers across chiplets and an optimized networking stack, this architecture keeps cores busy and exposes concurrent I/O for \iouring{} to be effective.

\paragraph{How to tune it.}
Once the engine operates asynchronously with large batched transfers, tuning \iouring{} can reduce data movement and CPU overhead per I/O.
Registered buffers and zero-copy send/receive eliminate kernel-user copies, reducing memory bandwidth consumption per unit of network throughput by $\approx 2\times$. 

\begin{figure}[]
  \centering
  \captionsetup{aboveskip=0.0ex,belowskip=0.0ex}
  \captionsetup[subfigure]{aboveskip=0.0ex,belowskip=0.0ex}
    \includegraphics[width=0.99\linewidth]{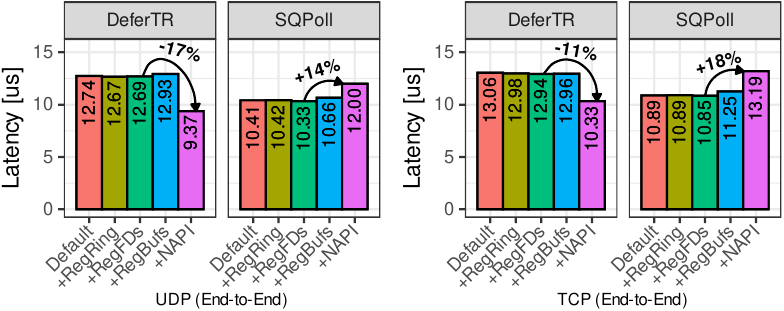}
    \caption{UDP and TCP latency ping-pong with 8~-byte messages and incrementally enabled optimizations. NAPI polling reduces latency for DeferTR, but increases it with SQPoll.} 
  \label{fig:netlatency}
\end{figure}

\subsection{Detailed Analysis of \iouring{}}
\label{subsec:net_benchmarks}
In this section, we focus on important aspects of \iouring{} for tuning network interfaces of database systems, which we could not cover before through targeted microbenchmarks.

\paragraph{Reducing latency with \iouring{}.}
Network latency is crucial for latency-sensitive protocols, such as transaction coordination or replication, where microsecond-level delays accumulate at scale.
We measure one-way and round-trip latencies for TCP and UDP using 8-byte messages to capture the minimum cost of message delivery.
Both modes - deferred taskrun (DeferTR) and SQPoll - are evaluated, along with optimizations such as registered file descriptors, registered buffers, and NAPI (the networking counterpart of IOPoll).
As shown in \Cref{fig:netlatency}, SQPoll achieves lower latency than DeferTR, but the advantage disappears once NAPI is enabled.
\hl{Without NAPI, \texttt{SQPoll} mainly helps by removing submission syscalls. With NAPI, the remaining handoff cost of \texttt{SQPoll} can dominate for very small messages.}
DeferTR with NAPI yields the best latency, outperforming SQPoll, while registered buffers have a negligible impact for small messages and increase latency slightly.
When the NIC queue is pinned to a remote chiplet, cross-chiplet interrupt handling increases latency by 14~\% for UDP and 21~\% for TCP.
With NAPI enabled, however, remote queues cause only a negligible increase of less than 1~\%.
For reference, a DPDK-based implementation reaches 7~µs, providing a lower bound for userspace networking.

\begin{figure*}[]
  \centering
  \captionsetup{aboveskip=0.0ex,belowskip=0.0ex}
  \captionsetup[subfigure]{aboveskip=0.0ex,belowskip=0.0ex}
    \includegraphics[width=0.99\linewidth]{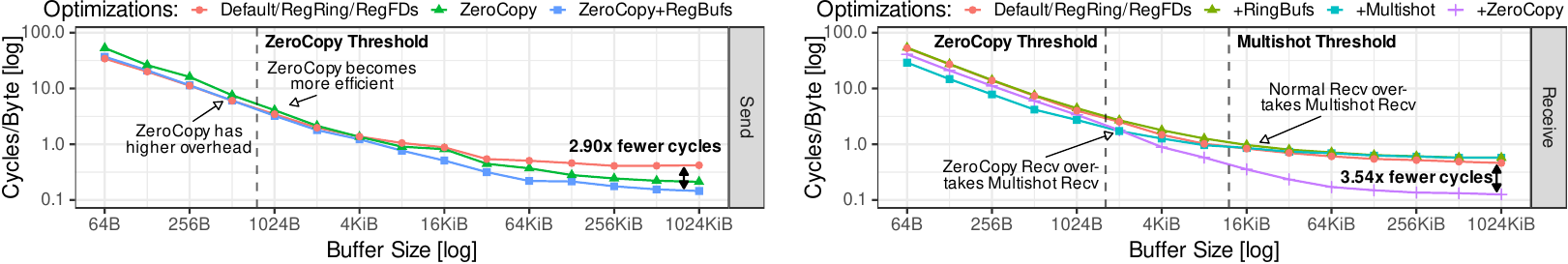}
    \caption{Impact of incremental \iouring{} optimizations on the cycle cost for a single TCP connection. The best-performing configuration depends on the shown thresholds. Registered file descriptors offer minimal benefit and are therefore omitted.} 
  \label{fig:zerocopythreshold}
\end{figure*}

\paragraph{Send path optimizations.}
\label{sec:net-zcthreshold}
As shown in \Cref{fig:netlatency}, registered buffers have little effect for very small messages, in contrast to their positive impact in the storage setting (\Cref{sec:storage_reg_buffers}).
To better understand which optimizations are most effective for different database workloads, we vary the transfer size and report the effective cycle cost per transmitted byte.
\Cref{fig:zerocopythreshold} (left) shows a clear threshold around 1~KiB: below this size, zero-copy send performs worse than plain \iouring{} due to buffer-management overheads, whereas for larger messages registered buffers amortize this cost and consistently reduce per-byte CPU time.
Registering rings and file descriptors yields only marginal improvements but introduces no observable drawbacks.
Overall, zero-copy with registered buffers is the most efficient configuration for large messages, achieving up to 3.5$\times$ fewer cycles per transmitted byte than default \iouring{}.

\paragraph{Receive path optimizations.}
We observe analogous thresholds for the receive path (\Cref{fig:zerocopythreshold}, right).
Multishot receive operations repeatedly generate completions from a single submission and are most efficient for workloads with small messages.
Once message sizes exceed roughly 1~KiB, zero-copy receive becomes more efficient, and for very large messages (e.g., 13~KiB and above) even the normal single-shot receive path outperforms multishot due to reduced per-message overheads.
\iouring{} can also draw receive buffers from a kernel-managed pool (RingBufs), but these perform worse than user-supplied buffers and are only useful in multishot scenarios.
Exact thresholds vary with capabilities and available offloads of the NIC, but the overall pattern remains consistent.

\paragraph{Optimizing kernel execution for sockets.}
Recall how \iouring{} executes \emph{task\_work} inside the kernel.
By default, it first attempts to complete an I/O operation inline in non-blocking mode and falls back to internal polling only if the call returns \texttt{-EAGAIN} (see \Cref{fig:iouring_work}).
For socket operations, this speculative attempt can be wasteful when the application already knows the socket is empty (for receive) or full (for send), causing unnecessary kernel work.
Such cases commonly occur in RPC-style communication, where the response is expected only after the request.
To handle this efficiently, \iouring{} provides the \ttsafe{RECVSEND_POLL_FIRST} flag, which skips the speculative attempt and directly uses the poll set.
Using PollFirst reduces the number of instructions executed and kernel work, resulting in up to 1.5$\times$ reduction in CPU cycles spent.


\section{Insights for System Builders}
\label{sec:insights}
In our study, we focused on three research questions: \emph{when to use \iouring{}}, \emph{how to integrate it}, and \emph{how to tune it}.
We summarize these insights as actionable guidelines for engineers, and then validate their practicality by applying them to enhance PostgreSQL's. 

\subsection{Guidelines}
\label{sec:guidelines}

Based on our system case studies, we draw four practical guidelines for using \iouring{} effectively in database systems:

\noindent
(1) \textbf{Determine if I/O is a system bottleneck.}
When I/O accounts for only a small fraction of execution time, as in CPU-bound or cache-resident workloads, potential \iouring{} gains are limited.
Our use cases in \Cref{sec:storage-intensive} and \Cref{sec:network-intensive_systems} show that \iouring{} is most effective when it reduces or amortizes the CPU cost of I/O operations, or when it lowers the memory bandwidth consumption.
As demonstrated in the buffer manager case study, simple latency or cycle models help to model such bottlenecks.

\noindent
(2) \textbf{Align the architecture with \iouring{} capabilities.}
\iouring{} enables asynchronous execution, system call batching, and a unified interface for storage and network I/O.
Our buffer manager (\Cref{subsec:buffer-manager}) shows that architectural changes, such as overlapping I/O and computation via asynchronous execution or amortizing per-I/O cost through batching, can yield large improvements.
The network shuffle (\Cref{subsec:shuffle}) shows how applications can use the \emph{ring-per-thread} architecture to scale beyond a single core. 

\noindent
(3) \textbf{Choose and tune the execution mode deliberately.}
The recommended \iouring{} configuration uses DeferTR with \emph{single-issuer} for predictable task execution and controlled completion reaping (\Cref{sec:background}).
SQPoll can improve performance when dedicating a polling core is amortized, and latency or IOPS targets justify the additional CPU cost.
Falling back to \textit{io\_workers} should be avoided by ensuring that all operations, including \texttt{fsync} and large I/Os, can execute fully asynchronously (cf.\ \Cref{fig:ssd_read_bs}).
Tuning the execution mode to the underlying hardware helps avoid costly effects, such as cross-chiplet traffic or non-local interrupt handling (\Cref{sec:shuffle_impl}).

\noindent
(4) \textbf{Use \iouring{} optimizations.}
\begin{highlight}
\iouring{} offers a range of general, storage-specific, and networking-specific optimizations for I/O-intensive systems.
Carefully selecting these optimizations can reduce the per-I/O cycle cost.
Some optimizations, such as registered FDs or fixed buffers for 4~KiB page-aligned storage I/O, do not negatively impact performance (see \Cref{sec:storage_reg_buffers}).
However, mechanisms like zero-copy or multishot receive are only effective when payloads exceed device-specific thresholds (cf.\ \Cref{fig:zerocopythreshold}).
Other optimizations, such as NVMe passthrough or IOPoll, apply only when no filesystem or a compatible filesystem is used.
\end{highlight}


\subsection{Optimizing PostgreSQL using Guidelines}
\label{sec:postgres}



\hl{PostgreSQL recently added \iouring{} support in version~18 for asynchronous data and WAL access~\cite{postgres18}. We use this integration to show how the guidelines apply in a production-grade engine despite PostgreSQL's architectural constraints to improve performance of a cold table scan from disk.}

\begin{figure}[]
  \centering
  \captionsetup{aboveskip=0.0ex,belowskip=0.0ex}
  \captionsetup[subfigure]{aboveskip=0.0ex,belowskip=0.0ex}
    \includegraphics[width=0.99\linewidth]{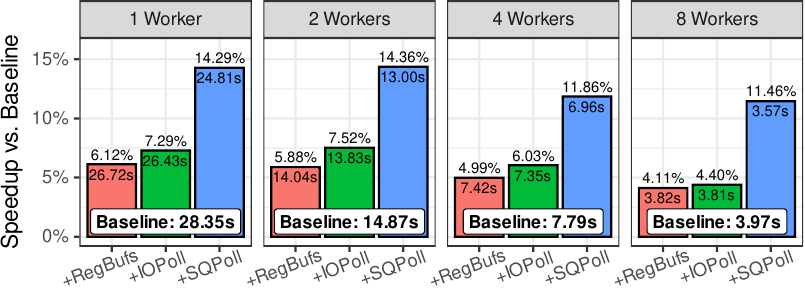}
      \caption{PostgreSQL speedup from \iouring{} optimizations. Sharing the SQPoll kernel thread between rings has negligible performance impact. Improvements remain limited by the filesystem and PostgreSQL's multi-process architecture.}
  \label{fig:postgres}
\end{figure}

\paragraph{GL~(1): Determine if I/O is a system bottleneck.}
\hl{Even without tuning \iouring{} parameters or enabling additional features, PostgreSQL's new \iouring{} backend achieves up to 3$\times$ higher performance in I/O-intensive workloads than the previous synchronous design based on blocking calls and OS readahead~\cite{postgres18}.
This confirms that I/O is a dominant bottleneck and provides the baseline for the improvements in \Cref{fig:postgres}, measured with 1--8 backend workers scanning a 32\,GiB cold table via direct I/O.}

\paragraph{GL~(2): Align the architecture with \iouring{} capabilities.}
PostgreSQL already partially aligns with guideline~(2) by overlapping computation and I/O through multiple asynchronous reads and writes.
\hl{However, its multi-process design allows backends to wait on I/O submitted through rings owned by other processes, so rings are not used by a single issuer and \texttt{DeferTR} cannot be applied directly.
A more \iouring{}-friendly design would use one ring per thread with exclusive ownership and no cross-process sharing.}

\paragraph{GL~(3): Choose and tune the execution mode deliberately.}
Given PostgreSQL's architecture, \texttt{DeferTR} cannot be used without substantial refactoring.
We therefore configure \iouring{} with \texttt{CoopTR} as the next-best alternative.
\hl{\texttt{CoopTR} avoids kernel-driven \emph{task\_work} preemptions while remaining compatible with PostgreSQL's process model.}
Still following guideline~(3), we extend the backend to support \texttt{SQPoll} by enabling the respective setup flag during ring creation and allowing multiple backend processes to share a single \texttt{SQPoll} kernel thread.
For WAL durability, PostgreSQL invokes \texttt{fsync()} directly (or via \ttsafe{O_DATASYNC}) rather than through \iouring{}, consistent with guideline~(3).
This avoids spawning \emph{io\_workers} for blocking \texttt{fsync()} calls and keeps the critical path fully asynchronous.

\paragraph{GL~(4): Use \iouring{} optimizations.}
\hl{Because PostgreSQL relies on filesystems for data storage, low-level optimizations such as NVMe passthrough are unavailable, and \texttt{IOPoll} depends on filesystem support.}
We apply guideline~(4) by enabling \iouring{} optimizations that fit PostgreSQL's access patterns and architectural constraints.
\hl{We register the entire buffer pool as fixed buffers, eliminating PostgreSQL's heuristic of enabling \texttt{IO\_ASYNC} after four outstanding I/Os. This matches our earlier findings for 8~KiB pages (\Cref{fig:ssd_read_bs}).}
In \Cref{fig:postgres}, fixed buffers alone provide a 4--6\% improvement.
Because we use ext4, we additionally enable \texttt{IOPoll}, improving performance up to 7.5\% over the baseline.
Combined with \texttt{SQPoll}, total throughput improves by 11--15\% over upstream PostgreSQL, despite the architectural constraints discussed above.

\paragraph{Summary.}
\hl{Applying the guidelines yields speedups even in a mature DBMS such as PostgreSQL, despite architectural and filesystem constraints that limit the applicability of several \iouring{} features.}

\section{Related Work}
\label{sec:related_work}

Although \iouring{} is a relatively recent addition to the Linux kernel, it has already been studied in various contexts, primarily focusing on storage I/O and its impact on data-intensive systems~\cite{DBLP:conf/btw/PestkaP25,DBLP:journals/corr/abs-2411-16254}.

\paragraph{Storage interfaces and NVMe systems.}
Initial systematic comparisons of \iouring{} with established interfaces such as \texttt{libaio} and SPDK were conducted using the \texttt{fio} benchmark to evaluate typical storage workloads \cite{DBLP:conf/systor/DidonaPIMT22}.
Ren and Trivedi extended this analysis for Intel Optane SSDs, characterizing microarchitectural behavior, instruction-level overheads, and Linux block I/O scheduler performance \cite{DBLP:conf/eurosys/RenT23}.
In the context of database systems, Haas et al.\ explored \iouring{} and other asynchronous I/O mechanisms for NVMe SSD arrays. 
They identified improvements in throughput and latency associated with various \iouring{} features \cite{DBLP:conf/cidr/HaasHL20,DBLP:journals/pvldb/HaasL23, haasManagingVeryLarge2025}.

\paragraph{Application-level and system integrations.}
Several works integrated \iouring{} into production systems to evaluate application-specific benefits.
Chen et al.\ applied \iouring{} to Redis~\cite{DBLP:conf/cloud2/ChenLLS24}, reporting significant reductions in overhead for medium and large payloads. 
Durner et al.\ used \iouring{} to accelerate cloud object storage access, achieving lower latency and higher throughput \cite{DBLP:journals/pvldb/DurnerL023}.
However, systematic analyses of network-specific aspects and end-to-end effects on distributed systems remain absent.

\paragraph{Security and advanced feature analyses.}
From a security perspective, He et al.\ proposed RingGuard~\cite{DBLP:conf/sigcomm/HeLZW23}, an eBPF-based framework that monitors and restricts \iouring{} operations to prevent kernel-level vulnerabilities.
It extends eBPF with new \iouring{}-specific hooks and verifier logic, enforcing safety policies at runtime while maintaining low overhead \cite{DBLP:conf/sigcomm/HeLZW23}.
The most comprehensive exploration of \iouring{} features to date is the work by Ingimarsson, who integrated basic functionality into RocksDB \cite{ingimarssonExploringPerformanceIo_uring2024}.
However, detailed evaluations of advanced \iouring{} features, such as registered buffers for zero-copy operations and linked requests to minimize system calls, are still lacking.

\begin{highlight}
\paragraph{Async I/O in Windows.}
Windows' \emph{IORing} API exposes shared submission and completion queues and supports batched submission, but the currently documented interface primarily targets file-oriented operations~\cite{msft-ioring,msft-ioringapi}.
Network I/O instead relies on the older \emph{Overlapped I/O} model, which supports both storage and networking but requires one system call per submitted request~\cite{msft-overlapped-io}.
Thus, unlike \iouring{}, current Windows interfaces do not provide a unified queue-based abstraction for both storage and network I/O.
\end{highlight}


\section{Conclusion and Outlook}
\label{sec:conclusion}

The modern Linux \iouring{} interface provides powerful mechanisms for asynchronous I/O, but achieving measurable gains requires more than simply enabling its features.
Our case studies show that performance depends on understanding system bottlenecks, integrating \iouring{} into the overall design, exploiting batching and asynchronous execution, and applying targeted optimizations to reduce CPU cost per I/O.
Focused microbenchmarks further highlight trade-offs such as zero-copy buffer-size effects and the importance of aligning polling strategies with hardware and workload characteristics.
From this analysis, we derived practical guidelines for I/O-intensive systems and validated them through an integration into PostgreSQL, improving table-scan throughput by 11--15\% over its baseline \iouring{} implementation.
As \iouring{} continues to evolve, adopting it today makes systems future-ready by enabling them to benefit from new capabilities such as zero-copy receive without architectural redesign.

\begin{acks}

This research was partially funded by the LOEWE Spitzenprofessur of the state of Hesse (III 5-519/05.00.003-(0005)), and by the Deutsche Forschungsgemeinschaft (DFG, German Research Foundation) under Germany’s Excellence Strategy (EXC3057/1 “Reasonable Artificial Intelligence”, Project No. 533677015).

\includegraphics[width=2em]{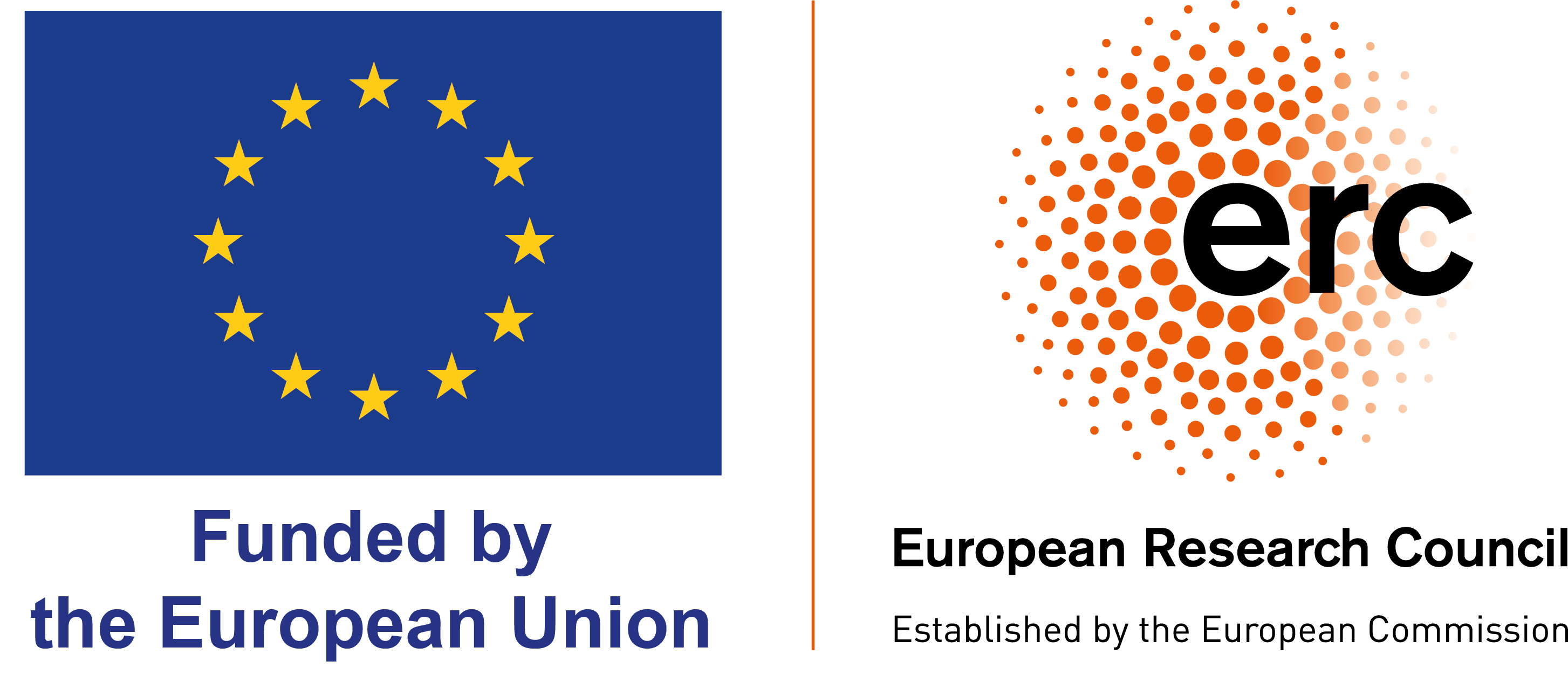} Funded/Co-funded by the European Union (ERC, CODAC, 101041375). Views and opinions expressed are however those of the author(s) only and do not necessarily reflect those of the European Union or the European Research Council. Neither the European Union nor the granting authority can be held responsible for them.

We thank Jens Axboe and the Linux kernel community for their valuable feedback and support regarding \iouring{}.
We also thank DFKI Darmstadt and hessian.AI for their support.
\end{acks}



\bibliographystyle{ACM-Reference-Format}
\bibliography{citation.bib}

\end{document}